\documentclass[11pt]{article}
\usepackage{moriond,epsfig}

\def\MyPhoto{\psfig{figure=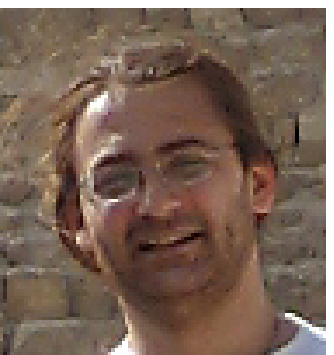,width=35mm}}

\def\Journal#1#2#3#4{{#1} {\bf #2}, #3 (#4)}

\def\PLB{{\em Phys. Lett.}  B}

\def\PRD{{\em Phys. Rev.} D}

\def\mco{\multicolumn}

\def\ra{\rightarrow}

\def\ko{K^0}

\def\be{\begin{equation}}
\def\ee{\end{equation}}
\def\bea{\begin{eqnarray}}
\def\eea{\end{eqnarray}}

\begin{document}
\vspace*{4cm}
\title{NEUTRINO DIPOLE MOMENTS AND SOLAR EXPERIMENTS}

\author{\underline{M. PICARIELLO}, B.C. CHAUHAN, C.R. DAS, FERNANDEZ-MELGAREJO, D. MONTANINO,\\
 J. PULIDO, E. TORRENTE-LUJAN}
\address{
Dipartimento di Fisica, Universit\`a del Salento - 
Via Arnesano, ex Collegio Fiorini, I-73100 Lecce, Italia\\
CFTP - Departamento de F\'\i sica, Instituto Superior T\'ecnico -
Lisboa, Portugal\\
Dep. de Fisica,  Univ. de Murcia - Murcia, Spain
}

\maketitle\abstracts{
First we investigate the possibility of detecting solar antineutrinos with
the KamLAND experiment.
Then we analyze the first Borexino data release to constrain the neutrino magnetic
moment. 
Finally we investigate the resonant spin flavour conversion of solar neutrinos to sterile ones,
a mechanism which is added to the well known LMA one. In this last condition, we show that the
data from all solar neutrino experiments except Borexino exhibit a clear preference for a
sizable magnetic field. We argue that the solar neutrino experiments are capable of tracing
the possible modulation of the solar magnetic field. In this way Borexino alone may play an
essential role although experimental redundancy from other experiments will be most important.
}

\section{Introduction}
Although the effort in solar neutrino investigation has decreased in recent years, several
intriguing questions in this area remain open.\cite{Raidal:2008jk}
Their clarification may lead to a better
knowledge of the neutrino intrinsic properties, the structure of the inner solar magnetic
field, or possibly both.
After having determined that the solar neutrino problem is essentially a
particle physics one and neutrinos oscillate,
the next step is to search 
for a possible new sub-dominant effects in the active solar neutrino flux and to 
investigate its low energy sector ($E<1-2~MeV$) which accounts for more than 
99\% of the total flux. These two issues in association with each other may 
lead to further surprises in neutrino physics, possibly the hint of a sizable 
magnetic moment. 
In fact it is still unclear for example whether the active solar 
neutrino flux varies in time \cite{Sturrock:2008qa} or why the SuperKamiokande 
energy spectrum appears to be flat.\cite{Fukuda:2002pe}

\section{Non standard neutrino interactions}

\subsection{KamLAND, solar antineutrinos and their magnetic moment}
\begin{figure}[h]
\rule{5cm}{0.2mm}\hfill\rule{5cm}{0.2mm}
\begin{center}
\vskip-0.5cm
\psfig{figure=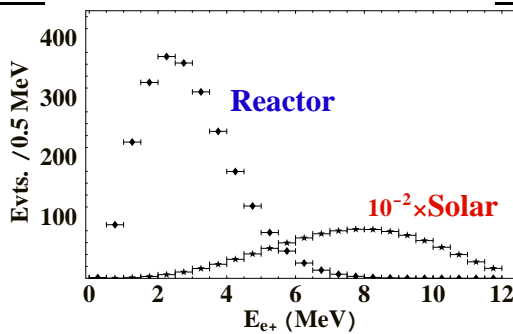,height=5.cm}
\vskip-0.5cm
\end{center}
\rule{5cm}{0.2mm}\hfill\rule{5cm}{0.2mm}
\caption{
The expected number of events if the solar neutrinos are converted into antineutrino with a factor $10^{-2}$.
\label{fig:KamLAND}}
\end{figure}

First we investigate the possibility to detect solar antineutrinos with
the KamLAND experiment.\cite{Aliani:2002pf}
These antineutrinos are, i.e., predicted by spin-flavor conversion
of solar neutrinos. The recent evidence from SNO shows that a) the neutrino
oscillates, only around 34\% of the initial solar neutrinos arrive at the Earth as electron
neutrinos and b) the conversion is mainly into active neutrinos, however a non $e$-$\mu$-$\tau$
component is allowed. The fraction of oscillation into  non-$\mu-\tau$ neutrinos 
is found to be $\cos^2\alpha= 0.08^{+0.20}_{-0.40}$. 
 This residual flux could include sterile neutrinos and/or the antineutrinos
of the active flavors.
KamLAND is potentially sensitive to antineutrinos derived from solar ${}^8$B neutrinos.
We report in fig. \ref{fig:KamLAND} the expected events at KamLAND
compared with a solar antineutrino flux $10^{-2}$ times the solar neutrino flux.

\noindent
KamLAND put strict limits on the flux
of solar antineutrinos, 
$\Phi( {}^8 B)< 1.0\times 10^4\ cm^{-2}\ s^{-1}$,
more than one order of magnitude smaller than existing limits,
and on their appearance probability  $P<0.15\%$ (95\% CL)
after 3 years of operation. Assuming a concrete model for 
antineutrino production by spin-flavor precession, this upper bound
implies an upper limit on the product of the intrinsic neutrino magnetic 
moment and the value of the solar magnetic field $\mu B< 10^{-21}$ MeV (95\% CL).
For $B\sim 10-100$ kG, 
we would have  $\mu <  10^{-11}-10^{-12}\ \mu_B$ (95\% CL).

\subsection{Three neutrinos: Limit from Borexino $|\mu_\nu| < 0.84 \times 10^{-10} \mu_B$}
Then we analyze the first Borexino data release to constrain the neutrino magnetic
moment.\cite{Montanino:2008hu}
The analysis is performed analyzing the spectrum of the recoil electron energy. Since the
leading contribution to this spectrum comes from the monoenergetic solar $^7Be$ neutrinos,
the shape of the spectrum is almost independent from the energy dependence of the
oscillation probability. The other contribution to the spectral shape is due to the internal
background of the detector.
The obtained limits are better than the one obtained for $SK-I$ global analysis
  $|\mu_\nu| < 3.6\times 10^{-10} \mu_B$,
and the combined analysis of the Kamiokande-Clorine experiments
 $|\mu_\nu| < 5.4\times 10^{-10} \mu_B$.
It is comparable with the combined analysis from other solar neutrino experiments, 
$|\mu_\nu| < 1.5\times 10^{-10} \mu_B$ at 90\% CL (SSM-GS98),
and with the Super Kamiokande total rate analysis,
$|\mu_\nu| < 2.1 \times 10^{-10} \mu_B$ at 90\% CL (SSM-AGS05).
It is competitive with respect to the direct limits from reactors
(i.e. $|\mu_\nu| < 1.0 \times 10^{-10} \mu_B$ at 90\% CL in MuNu,
$|\mu_\nu| < 0.58\times 10^{-10} \mu_B$ at 90\% CL in GEMMA experiment).
Moreover our result is independent on the solar standard model.
For the single transition magnetic moment we get
$|\mu_{\nu_\mu}| < 1.5\times 10^{-10} \mu_B$
(to be compared with the PDG value $< 6.8 \times 10^{-10} \mu_B$), and
$|\mu_{\nu_\tau}| < 1.9 \times 10^{-10} \mu_B$ (PDG quote $< 3900 \times 10^{-10} \mu_B$).

\section{Light sterile neutrinos and spin flavor precession}
\begin{figure}[h]
\rule{5cm}{0.2mm}\hfill\rule{5cm}{0.2mm}
\begin{center}
\vskip-0.2cm
\psfig{figure=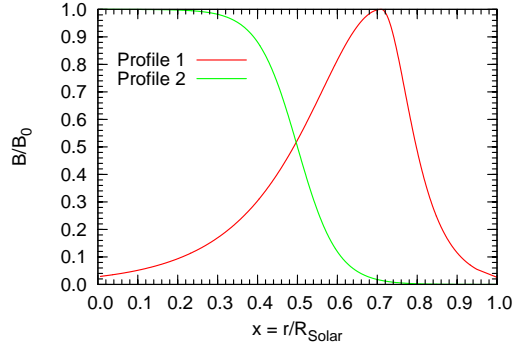,height=5cm}
\vskip-0.8cm
\end{center}
\rule{5cm}{0.2mm}\hfill\rule{5cm}{0.2mm}
\caption{
The two solar field profiles 
 normalized to their peak field values
as a function of the 
solar radius.
\label{fig:profiles}}
\end{figure}

As far as the solar magnetic field is concerned, solar physics provides
very limited knowledge on its magnitude and shape.
Given the above mentioned uncertainties we consider the two following plausible 
profiles which are approximately complementary to each other (see fig. \ref{fig:profiles})

{\it Profile 1}
\be
B=\frac{B_0}{\cosh[6(x-0.71)]} \quad 0<x<0.71\,,
\quad\quad
B=\frac{B_0}{\cosh[15(x-0.71)]} \quad 0.71<x<1
\ee

{\it Profile 2}
\be
B=\frac{B_0}{1+\exp[10(2x-1)]} \quad 0<x<1,
\ee

\noindent
Profile 1 has a peak $B_0$ at the bottom of the convection zone, for fractional 
solar radius $x\simeq 0.71$, its physical motivation being the 
large gradient of angular velocity over this range. It should 
not exceed 300 kG at this depth and 20 kG at 4-5\% depth, hence its fast decrease 
along the convection zone. Profile 2 is of the Wood-Saxon type, 
being maximal at the solar centre. In this case the peak field $B_0$ could be as 
large as a few MG.

The number of events in a bin with energy range $[T_e^{in},T_e^{fin}]$
is
\begin{eqnarray}
N_{T_e^{in}}^{T_e^{fin}}&=&\sum_{\phi=\nu\,fluxes}Q_0 N_{\Phi_{\phi}}
\int_{E_\nu=E_\nu^{min}}^{E_\nu^{max}} d\,E_\nu \Phi^\phi (E_\nu)
\int_{E_e=E_e^{min}}^{E_e^{max}} d\,E_e\, \tilde{R}(E_e,T_e^{in},T_e^{fin},\rho)
\nonumber\\&&\quad\quad
\sum_{\nu_x=\nu_e,\nu_\mu,\nu_\tau}\left\{P(E_\nu)_{\nu_e\rightarrow\nu_x}
\frac{d\sigma^{\nu_x}(E_e,E_\nu)}{d\,E_e}\right\}
\end{eqnarray}
where $Q_0$ take into account the size of the detector, $N_{\Phi_\phi}$ is the normalization of the neutrino flux $\phi$, $E_\nu$ is the neutrino energy, $E_e$ is the electron energy, $\Phi_\phi$ is the neutrino spectrum of the flux $\phi$, $\tilde R$ is the resolution function of the detector
and depend on the observed electron energy range and the detector properties $\rho$, $P(E_\nu)$ is the conversion/survival probability, and $d\sigma$ is
the differential neutrino-electron cross section.
For the statistical analysis of all solar data (except Borexino) we used the 
standard $\chi^2$ definition
\be
\chi^2=\sum_{ii^\prime}(R_j^{th}-R_j^{exp})(\frac{1}{\sigma^2})_{ji}(R_i^{th}-R_i^{exp})
\ee
where indices i,j run over solar neutrino experiments and the error matrix includes the cross section, the astrophysical and the experimental uncertainties.

\subsection{Two gallium data sets, spin flavour precession and KamLAND}
Although a lot of effort has been devoted to examining the
possible time modulation of the neutrino flux, 
this question remains largely unsettled. The claim made in the
early days of a possible anticorrelation 
of the Homestake event rate with sunspot activity remained 
unproven, as no sufficient evidence was found in its support. More recently the 
Stanford Group has been claiming the existence of two peaks \cite{Sturrock:2008qa} 
in the Gallium data at 55-70 SNU and 105-115 SNU. Moreover, Gallium
experiments, which have been running since 1990-91 and whose event rates 
are mainly due to $pp$ and $^7Be$ neutrinos (55\% and 25\% respectively), also
seem to show a flux decrease from their start until 2003. 
These data are hardly consistent with a constant value and exhibit a discrepancy of 
2.4$\sigma$ between the averages of the 1991-97 and 1998-03 periods (see table I). No 
other experiment sees such variations and none is sensitive to low energy neutrinos with 
the exception of Homestake whose rate contains only 14\% of $^7 Be$. Hence this fact 
opens the possibility that low energy neutrinos may undergo a time modulation partially 
hidden in the Gallium data which may be directly connected in some non obvious way to 
solar activity. Hence also the prime importance 
of the low energy sector investigation. To this end, in the near future, two experiments, 
Borexino and KamLAND, will be monitoring the $^7 Be$ neutrinos.
\begin{table}[h]
\caption{Average rates for Ga experiments in SNU. Set (I) and (II) reefer to
the period 1991-97 and 1998-03.}
\begin{center}
\begin{tabular}{|c|cc|} \hline
Period &  1991-97 (I) & 1998-03 (II) \\ \hline
SAGE+Ga/GNO & $77.8\pm 5.0$ & $63.3\pm 3.6$ \\
Ga/GNO only & $77.5\pm 7.7$ & $62.9\pm 6.0$ \\
SAGE only    & $79.2\pm 8.6$ & $63.9\pm 5.0$  \\ \hline
\end{tabular}
\end{center}
\end{table}

In a situation with light sterile neutrinos and spin flavor precession,
we reexamine the possibility of a time modulation of the low energy solar neutrino 
flux.\cite{Chauhan:2006yd} We perform two separate fits to the solar neutrino data, one 
corresponding to 'high' and the other to 'low' Ga data, associated with low
and high solar activity respectively. We therefore consider an alternative to 
the conventional solar+KamLAND fitting, which allows one to explore the much wider 
range of the $\theta_{12}$ angle permitted by the KamLAND fitting alone. We 
find a solution with parameters $\Delta m^2_{21}=8.2\times 10^{-5} eV^2, 
tan^{2}\theta=0.31$ in which the 'high' and the 'low' Ga rates lie far apart 
and are close to their central values and is of comparable quality to the global 
best fit, where these rates lie much closer to each other. This is an indication 
that the best fit in which all solar and KamLAND data are used is not a good 
measure of the separation of the two Ga data sets, as the information from the low 
energy neutrino modulation is dissimulated in the wealth of data. Furthermore 
for the parameter set proposed one obtains an equally good fit to the KamLAND 
energy spectrum and an even better fit than the 'conventional' LMA one for the 
reactor antineutrino survival probability as measured by KamLAND.
\begin{table}[h]
{\caption{Best fits to data sets , and LMA best fit. For data set 91-97 only Ga, Cl and Kamiokande data were available and for set 98-03 all SK and SNO data were available but not Cl.\,\, In set 98-03 only the Ga rate contri\-bu\-tes to $\chi^2_{rates}$.
Units are SNU for Ga and Cl and $10^6\ cm^{-2} s^{-1}$ for SK and SNO. Here $\Delta m^2_{01}=0.65\times 10^{-7} eV^2$.}}
\begin{center}
\begin{tabular}{|c|cccccccccc|} \hline
  & Ga & Cl & K (SK) & $\rm{SNO_{NC}}$ & $\rm{SNO_{CC}}$ & $\rm{SNO_{ES}}$ &
$\!\!\chi^2_{rates}\!\!$ & $\chi^2_{{SK}_{sp}}$ & $\chi^2_{{SNO}_{gl}}$ & $\chi^2_{KL}$\\ \hline 
Set (I)  & 71.7 &  2.66 &  2.29 &  &  &  & 3.09 &  &  & 15.3 \\ 
Set (II)  & 69.6 &   &  2.18 & 5.53 & 1.54 & 2.16 & 2.28 & 44.6 & 45.8 
& 15.3 \\ \hline  
LMA  & 64.8 &  2.74 &  2.30 & 5.10 & 1.75 & 2.28 & 0.95 & 45.7 & 43.1 & 14.5 \\ \hline
\end{tabular}
\end{center}
\end{table}
\subsection{SNO+: predictions from SSM and resonant spin flavour precession}
\begin{figure}[h]
\rule{5cm}{0.2mm}\hfill\rule{5cm}{0.2mm}
\begin{center}
\vskip -0.1cm
\psfig{figure=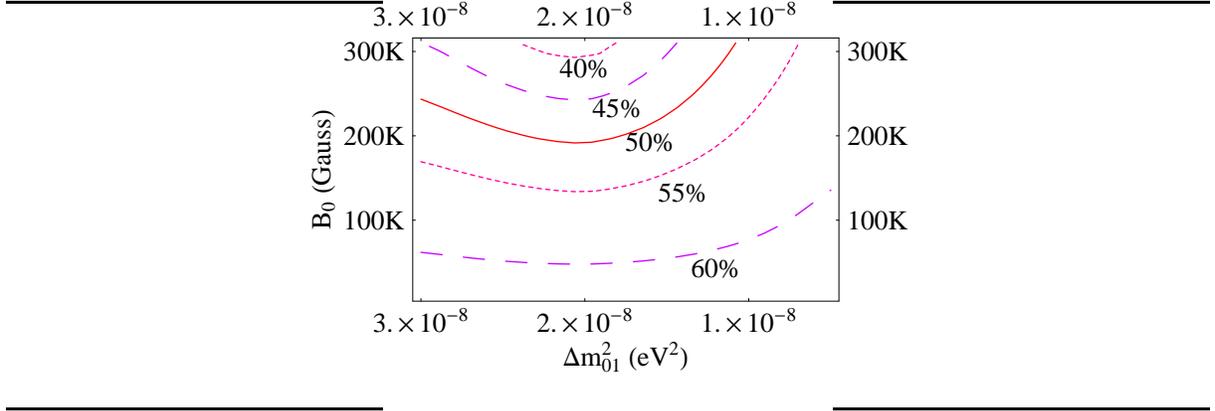,height=5cm}
\vskip -0.5cm
\end{center}
\rule{5cm}{0.2mm}\hfill\rule{5cm}{0.2mm}
\caption{
The expected rate reduction for the pep flux with respect to the non-oscillation case.
\label{fig:SNO}}
\end{figure}

One of the key questions that the SNO+ experiment will be able to address is the distinction
between the two classes of SSMs which are currently identified as corresponding to a high
and a low heavy element abundance. 
SNO+ will be able to accurately measure the pep and CNO fluxes. The former, largely
independent of solar models, will supply the survival probability at low energies, essential to
distinguish standard LMA from LMA+RSFP. Consequently SNO+ will be able to severely
constrain the RSFP interpretation, thus favoring LMA or vice-versa.\cite{Picariello:2007sw}
We report in fig. \ref{fig:SNO} the expected rate reduction for the pep flux with respect to the non-oscillation case, as a function of the peak value $B_0$ of the solar magnetic field (profile 1) and $\Delta m^2_{01}$.

\subsection{SuperKamiokande spectrum with light sterile neutrinos and spin flavour precession}
Whereas the Landau Zener approximation works well in the LMA resonance, this is not so 
for spin flavour precession, thus we resort to the numerical integration
of the evolution equations.\cite{Das:2009kw} We take several values of $\theta_{13}$ 
in the allowed range for both strong and weak solar fields. The model event 
rates for all solar neutrino experiments are evaluated and confronted with 
the data. Special emphasis is given to the SK energy spectrum 
\cite{Fukuda:2002pe} and the recent $^8B$ energy spectrum from the Borexino 
experiment.\cite{Collaboration:2008mr} We considered the two classes of solar 
field profiles. 
%
\begin{table}[h]
\caption{Peak field values (profile 1), $\sin \theta_{13}$, total rates 
(in SNU for Ga and Cl experiments, in $10^6 cm^{-2}s^{-1}$ for SK and SNO), and 
the corresponding $\chi^{2}$'s. 
It is seen that for a sizable field all fits improve with both profiles.}
\begin{center}
\begin{tabular}{|cc|cccccccccc|} \hline
$B_0$ & $\sin \theta_{13}$ & Ga & Cl & SK & $\!\!\rm{SNO_{NC}}\!\!$ & $\!\!\rm{SNO_{CC}}\!\!$ & $\!\!\rm{SNO_{ES}}\!\!$ &
$\!\!\chi^2_{rates}\!\!$ & $\chi^2_{{SK}_{sp}}$ & $\chi^2_{SNO}$ & $\chi^2_{gl}$\\ \hline
 & 0  & 67.2 & 2.99 & 2.51 & 5.62 & 1.90 & 2.49 & 0.07 & 42.7 & 57.2 & 99.9 \\
0 & 0.1 & 66.0 & 2.94 & 2.49 & 5.62 & 1.87 & 2.46 & 0.30 & 42.1 & 55.2 & 97.6 \\ 
 & 0.13 & 65.0 & 2.90 & 2.46 & 5.62 & 1.84 & 2.44 & 0.62 & 41.7 & 53.7 & 96.0 \\
\hline
Profile 1  & 0  & 66.4 & 2.82 & 2.32 & 5.37 & 1.76 & 2.31 & 0.20 & 37.6 & 46.0 & 83.8 \\
$140 (kG)$  & 0.1  & 65.3 & 2.77 & 2.29 & 5.37 & 1.73 & 2.28 & 0.53 & 37.9 & 44.9 & 83.3 \\
  & 0.13  & 64.3 & 2.72 & 2.27 & 5.37 & 1.70 & 2.25 & 0.95 & 38.4 & 44.1 & 83.4 \\
\hline
Profile 2  & 0  & 64.7 & 2.75 & 2.32 & 5.38 & 1.76 & 2.32 & 0.76 & 38.0 & 46.1 & 84.8 \\
$0.75 (MG)$  & 0.1  & 63.6 & 2.70 & 2.30 & 5.38 & 1.73 & 2.29 & 1.32 & 38.4 & 45.0 & 84.7 \\
  & 0.13  & 62.6 & 2.66 & 2.28 & 5.38 & 1.70 & 2.26 & 1.92 & 38.8 & 44.2 & 84.9 \\
\hline
\end{tabular}
\end{center}
\end{table}
\begin{figure}[h]
\rule{5cm}{0.2mm}\hfill\rule{5cm}{0.2mm}
\vskip -0.1cm
\begin{tabular}{lcr}
\psfig{figure=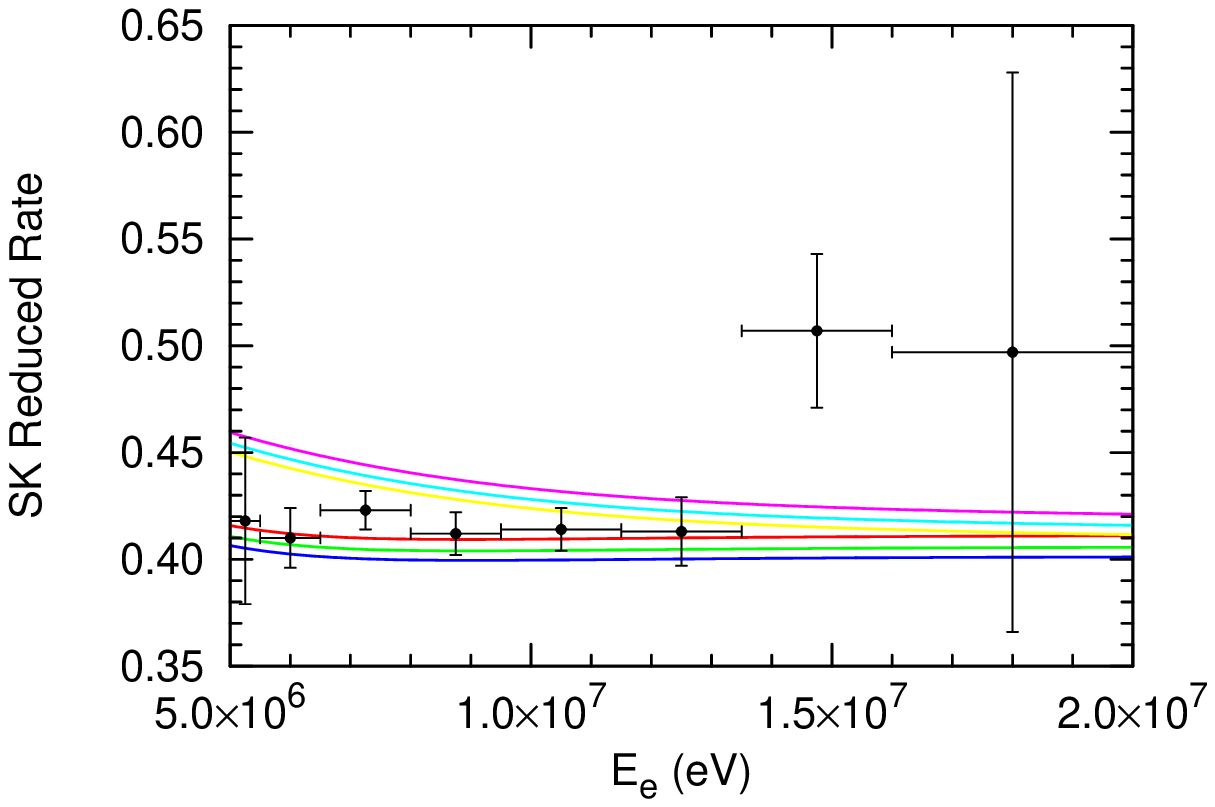,height=4.5cm}&$\quad\quad\quad$&
\psfig{figure=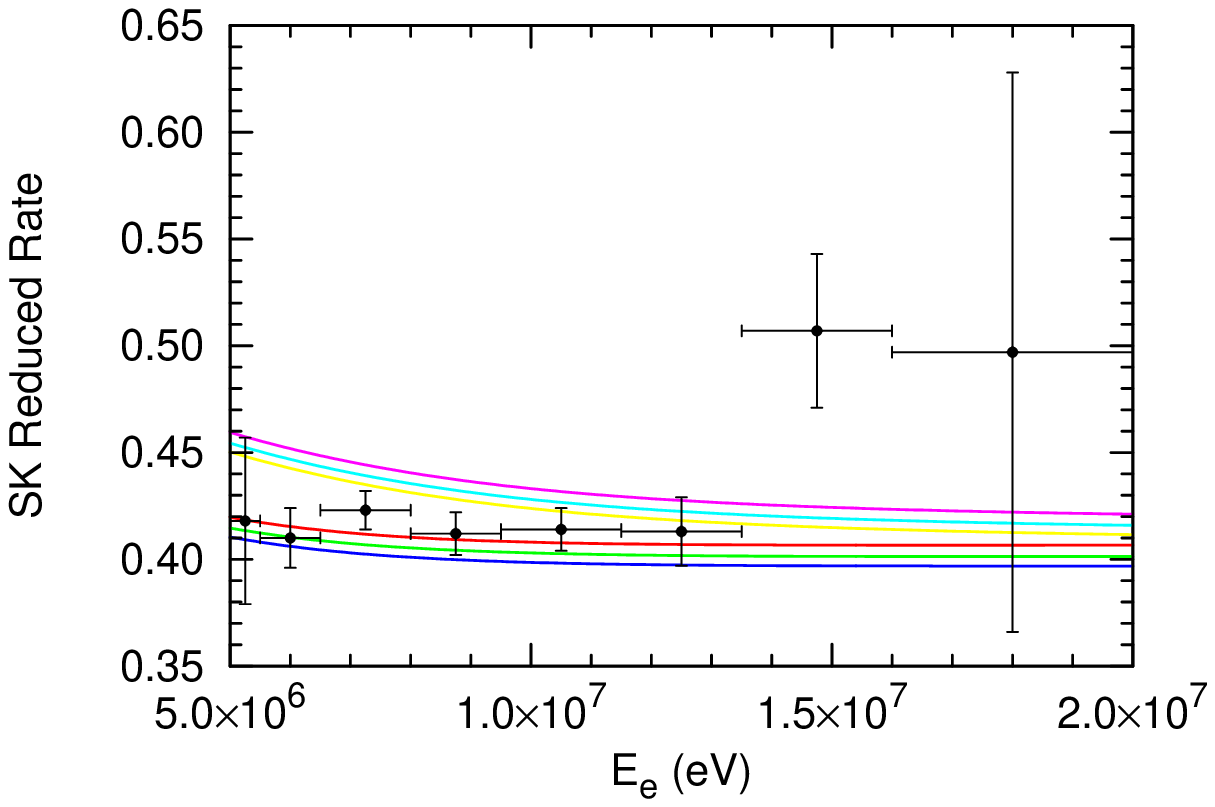,height=4.5cm}
\end{tabular}
\vskip -0.4cm
\rule{5cm}{0.2mm}\hfill\rule{5cm}{0.2mm}
\caption{
The SK spectrum: theoretical predictions and data points normalized
to BPS08(GS). The top three curves refer to $\sin\theta_{13} = 0, 0.1, 0.13$ from top
to bottom in the case of zero magnetic field, and the lower three curves refer to the same
values of $\sin\theta_{13}$ for a sizable field (Left: profile 1, $B_0 = 140\,kG$;
Right: profile 2).
There is a clear preference for a sizable field possibly related to solar activity, in
comparison to a vanishing one.
\label{fig:SK}}
\end{figure}
\\
Our numerical calculations are based on the updated central values for the best
fith in the LMA scenario for 
$\Delta m^2_{21}$, $\theta_{12}$, $\theta_{23}$, $\Delta m^2_{32}$ and we
use a neutrino transition moment between flavour states not larger than
$\mu_{\nu}=1.4\times 10^{-12}\mu_B$. As for $\theta_{13}$ we chose to investigate 
three cases: $\theta_{13}=0$, 0.1 and the central value, 0.13. The fits to all 
data, including rates and spectra (except for Borexino) improve once the 
magnetic field is introduced. As regards Borexino, the fit worsens in this 
case. 
In contrast, solar data alone show no clear preference for a 
vanishing or sizable $\theta_{13}$.
\section{Conclusions}
We studied several scenarios where non standard interactions, and in particular the neutrino magnetic
moment, as a mechanism which is added to the well known LMA one, can play a relevant
role in the solar neutrino physics.
The most promising one is when, in a 4 $\nu$ scenario, the transition magnetic moments from the $\nu_\mu$ and  $\nu_\tau$ to  $\nu_s$ play the dominant role in fixing the amount of active flavor suppression via 
the Resonant Spin Flavor Precession of Solar neutrinos to light sterile neutrino.
The data from all solar neutrino experiments except Borexino exhibit a clear preference
for a sizable magnetic field either in the convection zone or in the core and radiation zone.
We argue that the solar neutrino experiments are capable of tracing the possible modulation of the solar magnetic field.
Those monitoring the high energy neutrinos, namely the $^8B$ flux, appear to be sensitive to
a field modulation in the convection zone. 
Those monitoring the low energy fluxes will be sensitive to the second type of solar field
profiles only. 
In this way Borexino alone may play an essential role, since it examines both energy sectors,
although experimental redundancy from other experiments will be most important.

\section*{Acknowledgments}
One of us (M.P.) would like to thanks the organizers and the participants to the ``{\em XLIV$^{th}$ Rencontres de Moriond 2009 - EW 
for the very interesting and pleasant conference.


\section*{References}
\begin{thebibliography}{99}
\bibitem{Raidal:2008jk}
  M.~Raidal {\it et al.},
  Eur.\ Phys.\ J.\  C {\bf 57} (2008) 13.
\bibitem{Sturrock:2008qa}
  P.~A.~Sturrock,  arXiv:0810.2755 [astro-ph];
  L.~Pandola,
  Astropart.\ Phys.\  {\bf 22} (2004) 219.
\bibitem{Fukuda:2002pe}
  S.~Fukuda {\it et al.}  [Super-Kamiokande Collaboration],
  Phys.\ Lett.\  B {\bf 539} (2002) 179.
\bibitem{Collaboration:2008mr}
  [Borexino Collaboration], G. Bellini {\it et al.},
  arXiv:0808.2868 [astro-ph].
\bibitem{Aliani:2002pf}
  P.~Aliani, V.~Antonelli, M.~Picariello and E.~Torrente-Lujan,
  JHEP {\bf 0302} (2003) 025.
\bibitem{Montanino:2008hu}
  D.~Montanino, M.~Picariello and J.~Pulido,
  Phys.\ Rev.\ D {\bf 77} (2008) 093011.
\bibitem{Chauhan:2006yd}
  B.~C.~Chauhan, J.~Pulido and M.~Picariello,
  J.\ Phys.\ G {\bf 34} (2007) 1803.
\bibitem{Picariello:2007sw}
  M.~Picariello {\it et al.},
  JHEP {\bf 0711} (2007) 055.
\bibitem{Das:2009kw}
  C.R.~Das, J.~Pulido and M.~Picariello,
  Phys.\ Rev.\ D {\bf 79} (2009) 073010. 
\end{thebibliography}
\end{document}